\documentclass[12pt]{article}
\usepackage{graphicx,tabularx,epsfig}
\usepackage{color}
\usepackage{enumitem}
\usepackage{caption}
\usepackage{subcaption}
\usepackage{amsmath}
\usepackage{amssymb}
\usepackage{amsthm}
\usepackage{physics}
\usepackage{dsfont, mathtools}
\usepackage{psvectorian}
\usepackage{blkarray}
\usepackage{bm}
\usepackage{url}
\usepackage{setspace}

\usepackage{xcolor}
\colorlet{linkequation}{blue}

\usepackage[colorlinks = true,
            linkcolor = blue,
            urlcolor  = blue,
            citecolor = blue,
            anchorcolor = blue]{hyperref}

\usepackage{floatrow}

\newlength{\abstractwidth}
\renewcommand{\thefootnote}{\fnsymbol{footnote}}
\renewcommand{\thanks}[1]{\footnote{#1}} 
\newcommand{\starttext}{
\setcounter{footnote}{0}
\renewcommand{\thefootnote}{\arabic{footnote}}}

\makeatletter
\g@addto@macro\normalsize{%
  \setlength\abovedisplayskip{10pt}
  \setlength\belowdisplayskip{20pt}
  \setlength\abovedisplayshortskip{10pt}
  \setlength\belowdisplayshortskip{20pt}
}
\makeatother

\usepackage{color}

\renewcommand{\title}[1]{\vbox{\center\LARGE{#1}}\vspace{5mm}}
\renewcommand{\author}[1]{\vbox{\center#1}\vspace{5mm}}

\makeatletter
\newcommand{\lambdabar}{{\mathchoice
  {\smash@bar\textfont\displaystyle{0.25}{1.2}\lambda}
  {\smash@bar\textfont\textstyle{0.25}{1.2}\lambda}
  {\smash@bar\scriptfont\scriptstyle{0.25}{1.2}\lambda}
  {\smash@bar\scriptscriptfont\scriptscriptstyle{0.25}{1.2}\lambda}
}}
\newcommand{\smash@bar}[4]{%
  \smash{\rlap{\raisebox{-#3\fontdimen5#10}{$\m@th#2\mkern#4mu\mathchar'26$}}}%
}
\makeatother

\begin{document}

\singlespacing


  
\begin{titlepage}

\rightline{}
\bigskip
\bigskip\bigskip\bigskip\bigskip
\bigskip

\centerline{\Large \bf {Complexity and Momentum}}

\bigskip \noindent

\bigskip
\begin{center}
\bf  Leonard Susskind$^1$$^2$ and Ying Zhao$^3$  \rm

\bigskip

$^1$SITP, Stanford University, Stanford, CA 94305, USA \vskip0.4em
$^2$Google, Mountain View, CA 94043, USA\vskip0.4em
$^3$Institute for Advanced Study, Princeton, NJ 08540, USA

\end{center}

\bigskip \noindent
\begin{abstract}

Previous work has explored the connections between three concepts---operator size, complexity, and the bulk radial momentum of an infalling object---in the context of JT gravity and the SYK model. In this paper we investigate the higher dimensional generalizations of these connections.  We use a toy model to study the growth of an operator when perturbing the vacuum of a CFT. From circuit analysis we relate the operator growth to the rate of increase of complexity and check it by complexity-volume duality. We further give an empirical formula relating complexity and the bulk radial momentum that works from the time that the perturbation just comes in from the cutoff boundary, to after the scrambling time.

\end{abstract}

\end{titlepage}

\starttext \baselineskip=17.63pt \setcounter{footnote}{0}

{\hypersetup{hidelinks}
\tableofcontents
}

\section{Introduction}

In \cite{Susskind:2018tei} it was conjectured that  as an object falls toward a gravitating mass,  the increase of bulk radial momentum is related to the growth of  boundary operators. The relation was best understood in the case of $AdS_2$ gravity and SYK\cite{Brown:2018kvn}\cite{Qi:2018bje}. It was also pointed out that even in the absence of an explicit gravitating mass, vacuuum  $AdS_2$ itself has a gravitational field (due to the vacuum energy) that attracts objects to the center, and that that falling through empty $AdS_2$ is also accompanied by operator growth \cite{Susskind:2019ddc}.

 On the other hand, from the study of the $SL(2)$ symmetry generators in JT gravity\cite{Lin:2019qwu}, the momentum generator can be directly related to complexity by $2\pi P = \frac{d\mathcal{C}}{dt}$. The known relation between complexity and operator size then ties the three concepts together---momentum, size, and complexity \cite{Susskind:2019ddc}.
In this paper we will generalize this connection between operator growth, complexity, and radial momentum to higher dimensions. 



In the presence of a black hole, operator growth in the scrambling regime was well studied\cite{Sekino:2008he}\cite{Shenker:2013pqa}\cite{Stanford:2014jda}\cite{Susskind:2014jwa}\cite{Roberts:2014isa}. It can be analyzed by an epidemic toy model \cite{Susskind:2014jwa}. In this paper, we look at a different regime---the vacuum AdS regime---and consider the growth of operators size  when perturbing empty CFT. 

To define size, the first question one needs to answer is, what are the size-one objects? In SYK, the size-one objects are fermions.\footnote{The size needs to be renormalized at finite temperature. See \cite{Qi:2018bje}.} In the scrambling regime, the size-one objects are thermal scale quanta\cite{Sekino:2008he}. In the vacuum regime, we will see that the size-one objects are energy scale dependent. As the perturbation flows from UV to IR, the basic excitations carry lower and lower energy and the number of excitations increases. Size counts the number of these basic excitations. From the bulk point of view, the particle falls into a deeper and deeper radial location and gets a higher and higher momentum. We study this process by a toy model of gluon-spitting and see that the operator size grows linearly in time in this regime, and this parallels the linear growth of momentum. 

It is well known that operator growth is also related to the increase of complexity\cite{Susskind:2014jwa}\cite{Roberts:2014isa}: $\text{size} = \frac{d\mathcal{C}}{d\tau}$ where $\tau$ is the circuit time. In the scrambling regime the conversion factor between the circuit time and the boundary time is $\frac{2\pi}{\beta}$. In the vacuum regime the conversion factor between the circuit time and the boundary time is again energy scale dependent. Taking account of this factor one can show that complexity also grows linearly in time. From another direction, the change of complexity is related to the change of bulk volume after the perturbation\cite{Stanford:2014jda}. We compute the volume change right after perturbing the vacuum by a spherically symmetric perturbation and see that it indeed implies that the complexity grows linearly in time.

The relation between operator size and momentum near the Rindler region was pointed out in \cite{Susskind:2018tei}. The relation between operator size/complexity and momentum in $AdS_2$ throat was studied in \cite{Brown:2018kvn}\cite{Qi:2018bje}\cite{Susskind:2019ddc}\cite{Lin:2019qwu}. We propose a formula relating the radial momentum to the rate of increase of complexity: $\frac{\lambdabar}{2\pi}\frac{d\mathcal{C}}{dt} = 2\pi P$ where $\lambdabar$ is the dimensionless boundary wavelength of the excitation,
\begin{equation}
\lambdabar = \frac{\lambda}{2\pi l_{AdS}}.
\end{equation}

 This formula works in $AdS_2$ as well as higher dimensions, from the time that the perturbation  comes in from the cutoff boundary, until after the scrambling time. It is an empirical formula that is justified by the results. It would be  interesting to give a first principles derivation.

The rest of the paper is organized as follows. In section \ref{2D} we review the connections between operator growth, complexity increase, and particle momentum in $AdS_2$ gravity and SYK. In section \ref{operator_growth} we study the operator growth in spacetime dimension $D\geq 3$ and introduce a gluon-splitting toy model describing the growth of the operator when we perturb the vacuum of a CFT. In section \ref{complexity} we relate the operator growth to complexity increase and compare our toy model analysis with a CV calculation. In section \ref{size_momentum} we make a connection between the rate of increase of complexity and the bulk radial momentum. We check it in various examples. We conclude the paper in section \ref{conclusion} and point out some unanswered questions.

\section{Review of operator growth in SYK and infalling particle in $AdS_2$}
\label{2D}
The growth of a simple operator under time evolution in SYK was studied in \cite{Roberts:2018mnp}\cite{Qi:2018bje}. Starting from one fermion $\psi_1$, the average number of fermions making up the operator $\psi_1(t)$ increases as time increases. In \cite{Susskind:2018tei}\cite{Brown:2018kvn} it was pointed out the growth of the operator corresponds to the increase of the particle momentum as it falls in. More precisely, in $AdS_2$ geometry and SYK we have
\begin{align}
\label{AdS2_1}
s_{\beta}(\psi(t)) = P\tilde\beta
\end{align}
In the above equation, $s_{\beta}(\psi(t))$ represents the average number of fermions in the operator $\psi(t)$ in a state with temperature $\frac{1}{\beta}$ \cite{Qi:2018bje}. The operator $\psi(t)$ produces an infalling particle in the dual bulk geometry. On the right hand side of equation \eqref{AdS2_1}, $P$ is the radial momentum of the particle. $\tilde T = \frac{1}{\tilde\beta}$ is the local energy scale depending on the radial location of the particle\cite{Brown:2018kvn}.

From another point of view, one can look for the $SL(2)$ symmetry generators of $AdS_2$ in JT gravity\cite{Lin:2019qwu}\cite{Susskind:2019ddc}. The momentum generator was found to be 
\begin{align}
\label{AdS2_2}	
  P \sim N\mathcal{J}\frac{d L}{dt}
\end{align}
where $L$ is the length of the wormhole and $\mathcal{J}$ is the energy scale in SYK                         .  

In fact, \eqref{AdS2_1} and \eqref{AdS2_2} are related. From CV duality\cite{Stanford:2014jda}, one can rewrite \eqref{AdS2_2} as 
\begin{align}
    \label{AdS2_3}
	2\pi P = \frac{d\mathcal{C}}{dt}
\end{align}
From quantum circuit consideration, one can relate the operator size to the complexity of the perturbed state\cite{Susskind:2014jwa}\cite{Roberts:2014isa}. Then from \eqref{AdS2_3} we can recover \eqref{AdS2_1}. 
\begin{align*}
	2\pi P = \frac{d\mathcal{C}}{dt} = \frac{d\mathcal{C}}{d\tau}\frac{d\tau}{dt} = s\frac{2\pi}{\tilde\beta} 
\end{align*}
where $\tau$ is the circuit time. $\tilde\beta$ gives the transformation between circuit time and boundary time.

In what follows, we will generalize the above discussions to bulk dimensions $D\geq 3$.

\section{Operator growth in higher dimensions}
\label{operator_growth}

In this section we consider toy models of operator growth in higher dimension. In the presence of a black hole the scrambling regime can be modeled by an epidemic picture \cite{Susskind:2014jwa}\cite{Roberts:2014isa} in which  the system is modeled by $S$ qubits, and we perturb the system by throwing in $\delta S$ extra qubits. The number of qubits affected by these $\delta S$ extra qubits grows exponentially in time until saturation. In this section, we will study a different regime (perturbed vacuum) and fit the two regimes together.

\subsection{A toy model: gluon splitting picture}

The gluon-splitting model we will use is an extrapolation of  weakly coupled gauge theory into the strongly coupled region. For weak coupling, $g_{ym}<<1,$ the model could be derived from perturbation theory. Our assumption, which is not new, is that the concept of a gluon and its splitting into a cascade of lower energy gluons has a strongly coupled limit.

In order to be concrete we introduce a cutoff  into the CFT corresponding to a maximum value of the radial coordinate $r_c$. We then consider a local gauge invariant operator---for example a single trace operator---in the cutoff theory and smear it over space (but not time) so that the tranverse momentum that it injects into the bulk is small. Such an operator will inject an finite energy  
 $E$ into the bulk corresponding to a pair of oppositely moving  gluons in the boundary theory. The  wavelengths of the gluons is equal to the cutoff. We say that each high energy gluon has size one.

 Such an UV perturbation will flow to the IR. At step one, it splits into four gluons, each with energy $\frac{E}{4}$. At step two we have eight gluons, each with energy $\frac{E}{8}$. At step $n$, we will have $2\times 2^n$ gluons, each with energy $\frac{E}{2^{n+1}}$. Each step takes longer time as we flow to IR.  If step one takes time $\delta t$, step two will take time $2\delta t$, and step $n$ takes time $2^n\delta t$. 
 
 In perturbation theory the rate of gluon splitting is proportional to  $g_{ym}^2$. One would find   $$ f(g_{ym}) \delta t\sim \frac{l_{ads}^2}{r_c}$$
 where $f(g_{ym})$ is an expansion in powers of the coupling. Our model assumes that $f(g_{ym}) $ tends to a constant $\sim 1$ at large coupling, but that otherwise perturbative reasoning makes sense.

 If the process starts at time $t = 0$, step $n$ happens at time $2^n\delta t$. Notice that this discrete time step is essentially RG step\cite{Swingle:2009bg}\cite{Swingle:2012wq}. We see that the size grows exponentially in RG steps and linearly in time.

\begin{align}
	&s(n) = s_02^n  \label{size_discrete}\\
	&s(t) =s_0\frac{t}{\delta t} \label{size_time1}
\end{align}

where $s_0$ is the initial size. 

\bigskip \noindent

To summarize, our basic hypothesis is that the gluons are the objects of size one, and that the size of the growing operator is simply the number of gluons in the cascade. As time advances the number of gluons grows while the energy of each gluon decreases. The wavelength of the gluons grows accordingly.
And of course from UV-IR connection\cite{Susskind:1998dq}, lower energy in the boundary theory corresponds to deeper radial location in the bulk. \\

Next we compare the above toy model to $AdS$ bulk geometry. We consider $AdS_D$ in the following coordinates: 
\begin{equation}
\label{vaccum_AdS}
	ds^2 = -\left(\frac{r^2}{l^2}+1\right)dt^2+\frac{dr^2}{\frac{r^2}{l^2}+1}+r^2d\Omega_{D-2}^2
\end{equation}

The radial location $r$ corresponds to boundary excitations with dimensionless wavelength $\lambdabar(r) = \frac{2\pi l}{r}$.\footnote{Consider a sphere of radius $r$. The number of degrees of freedom inside such a sphere is $\frac{r^{D-2}}{l_p^{D-2}}\sim N^2 \frac{r^{D-2}}{l^{D-2}}$. The number of unit cells is given by $\frac{r^{D-2}}{l^{D-2}}$. So each cell has wavelength  $\sim\frac{l}{r}$. The factor of $2\pi$ is chosen here to simplify later expressions.} In other words, when the bulk particle is at radial location $r$, we count the number of gluons with energy $\frac{1}{ l\lambdabar(r)}$. The size is given by 
\begin{align*}
	s(r) = El\lambdabar(r)
\end{align*}

Say, we create some excitation with energy $E$ at cutoff surface with radius $r_c$. The initial size is given by $s_0 =\frac{2\pi El^2}{ r_c}$. When the particle reaches radius $r$, its size becomes
\begin{align}
\label{size_continuous}
	s(r) =  2\pi E\frac{l^2}{r} = s_0\frac{r_c}{r}
\end{align}

Alternatively, at large $r$, the radial proper distance $\rho$ from the surface with radius $r$ to the cutoff surface satisfies $\frac{r_c}{r} = e^{\frac{\rho}{l}}$. We can identify the radial proper distance $\rho$ with RG step:
\begin{align*}
	n\log 2\sim\frac{\rho}{l}.
\end{align*} 
Then \eqref{size_continuous} is the same as \eqref{size_discrete}:
\begin{align*}
	s(\rho) = s_0 e^{\frac{\rho}{l}} 
\end{align*}

We look at the time dependence. Consider a particle falling into AdS. Assume $\frac{r}{l}\gg 1$, the radial geodesic satisfies
\begin{align}
\label{geodesic}
	\frac{l^2}{r^2}=\frac{l^2}{r_c^2}+\frac{t^2}{l^2}
\end{align}
From the relation between radial location and boundary wavelength, \eqref{geodesic} can also be written as
\begin{equation}
\label{wavelength}
	\left(\frac{\lambdabar(r)}{2\pi}\right)^2 = \left(\frac{\lambdabar(r_c)}{2\pi}\right)^2+\frac{t^2}{l^2}
\end{equation}

 With $r_c\rightarrow \infty$, we have $\frac{\lambdabar(r)}{2\pi} = \frac{t}{l}$. The wave length of size-one object increases linearly in time. Then \eqref{size_continuous} can be written as
\begin{align}
\label{size_time2}
	s(t) = 2\pi Et
\end{align}
\eqref{size_time2} is the same with \eqref{size_time1} if we identify $\delta t=\frac{ s_0}{2\pi E} = \frac{l^2}{r_c}$.

\subsubsection{The radius of gyration}

We look at these size-one objects from another perspective. In \cite{Coleman:1977yb} Coleman and Smarr defined the following quantity. Consider non-singular solution of classical gauge field theory. Let $R^2 = \sum_{i}^3x^ix^i$.
\begin{align*}
	\bar R(t)^2 \equiv \int d^3 x R^2 T^{00}(t) /E
\end{align*}
where $E = \int d^3x T^{00}$. $\bar R(t)$ is called the radius of gyration. Assume $T$ is conserved and traceless, one has
\begin{align}
\label{gyration}
	\bar R(t)^2 = R_0^2+t^2.
\end{align}
The authors discussed classical field theory. The properties they used are: 1. The stress-energy tensor falls off fast enough at infinity. 2. The stress energy tensor is conserved: $\partial_{\mu}T^{\mu\nu} = 0$. 3. Scale invariance: $T^{\mu}_{\mu} = 0$. As these properties can be generalized to quantum theory for the vacuum of CFT, their conclusion holds in the quantum theory if we consider expectation value of operators. Note that these are also the assumptions we used in the gluon-splitting toy model. 

From its definition, $\bar R(t)$ is the typical wavelength of the excitation. We also see that \eqref{gyration} is exactly the same with \eqref{geodesic}\eqref{wavelength} when we identify $\frac{\bar R}{l} = \frac{\lambdabar}{2\pi} = \frac{ l}{r}$. 

If we start from a local excitation from the boundary, $R_0 \approx 0$ and $\bar R(t)\approx t$. The wavelength of size-one object increases linearly in time as discussed in the gluon-splitting model.  

\subsection{Perturbing finite temperature state}
\label{finite_temperature}
\begin{figure}[H] 
 \begin{center}                      
      \includegraphics[width=2.6in]{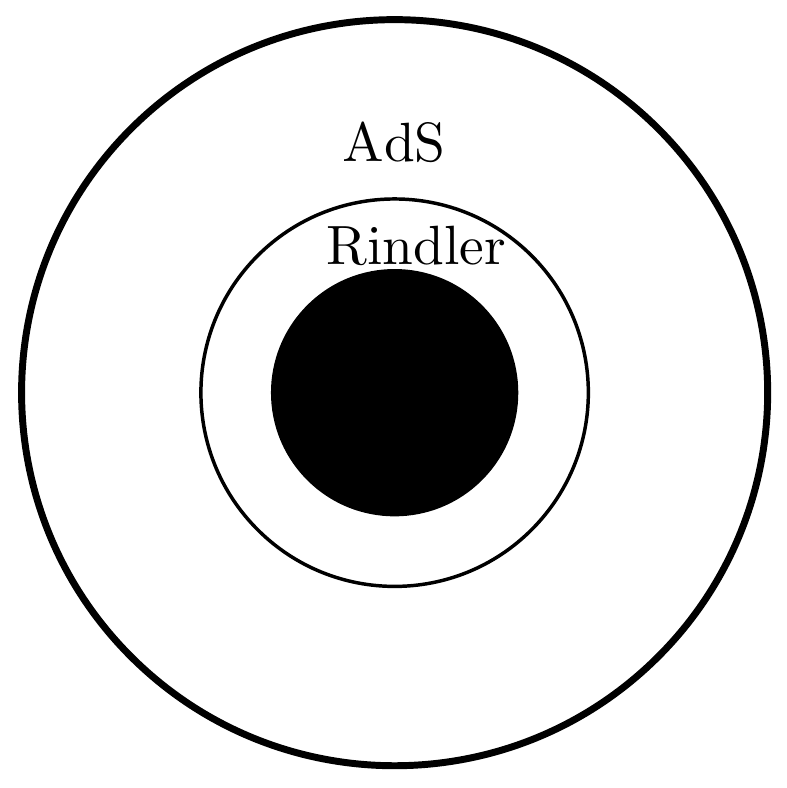}
      \caption{}
  \label{large_BH}
  \end{center}
\end{figure}

So far our discussion was in the vacuum state of a CFT. In this section we look at a finite temperature state. The dual geometry contains a black hole (Figure \ref{large_BH}).
\begin{align*}
	ds^2 = -f(r)dt^2+\frac{dr^2}{f(r)}+r^2d\Omega_{D-2}^2
\end{align*}
where $f(r) = \frac{r^2}{l^2}-\frac{2m}{r^{D-3}}+1$.

We take $D=3$ and look at BTZ black hole where $f(r) = \frac{r^2-r_h^2}{l^2}$. A timelike geodesic released at time zero from some cutoff radius $r = r_c$ satisfies
\begin{equation}
\label{geodesic_BTZ}
	\frac{l^2}{r^2} = \frac{l^2}{r_c^2}+\left(\frac{l^2}{r_h^2}-\frac{l^2}{r_c^2}\right)\tanh^2\left(\frac{r_h}{l^2}t\right)\approx \frac{l^2}{r_h^2}\tanh^2\left(\frac{r_h}{l^2}t\right)
\end{equation}
$r_h$ is the horizon radius. If we compare \eqref{geodesic_BTZ} with \eqref{geodesic}, we see that unlike the vacuum case, the energy scale stops decreasing once reaching temperature scale: $\frac{\lambdabar}{2\pi}=\frac{l}{r}\approx \frac{l}{r_h} = \frac{\beta}{2\pi l}$.

At energy scales larger than the temperature, the particle sees vacuum and earlier discussion applies. The size in vacuum is given by
\begin{align*}
	&s(r) =  El\lambdabar(r) = 2\pi E\frac{l^2}{r}\\
	&s(t) = 2\pi E\frac{l^2}{r_h}\tanh(\frac{r_h}{l^2}t)\approx 2\pi Et,\ \ t<\frac{l^2}{r_h}
\end{align*}
As $r$ decreases, $\lambdabar(r)$ increases and the particle size increases linearly in time. This process continues until the energy scale equals the temperature, i.e., $\frac{\lambdabar(r)}{2\pi}= \frac{l}{r_h} =\frac{\beta}{2\pi l}$, at which point the size becomes
\begin{align}
\label{transition}
	s_1 = E\beta = \delta S 
\end{align} 
At this point, the exciation is made of $\delta S$ quanta with thermal scale energy. Note that from \eqref{transition} $\delta S$ is nothing but the increase of black hole entropy due to this perturbation. After this, we enter the scrambling regime and the operator size increases exponentially in time until saturation \cite{Susskind:2014jwa}\cite{Roberts:2014isa}. 
\begin{equation}
	s(t) = \frac{\delta Se^{\frac{2\pi}{\beta}\left(t-\frac{l^2}{r_h}\right)}}{1+\frac{\delta S}{S}e^{\frac{2\pi}{\beta}(t-\frac{l^2}{r_h})}} \approx \begin{cases}
		\delta S e^{\frac{2\pi}{\beta}(t-\frac{l^2}{r_h})} & \frac{l^2}{r_h}<t<\frac{\beta}{2\pi}\log\frac{S}{\delta S}\\
		S & t>\frac{\beta}{2\pi}\log\frac{S}{\delta S}
	\end{cases}
\end{equation}
In scrambling regime the energy scale no longer changes.

\section{Operator growth and complexity increase}
\label{complexity}

\subsection{Quantum circuit consideration}
\label{size_estimation}

In \cite{Susskind:2014jwa}\cite{Roberts:2014isa} it was argued that the operator size growth is related to complexity increase. Roughly speaking, in each circuit time step, the increase of complexity equals to the size of the operator at that instant: $\frac{d\mathcal{C}}{d\tau} = s(\tau)$ where $\tau$ is circuit time. 

Note that the circuit time $\tau$ is not the same as boundary time $t$. In vacuum regime when a particle falls into empty AdS, the circuit time is the RG step. It corresponds to the radial proper distance $\rho$. If we want to know what is $\frac{d\mathcal{C}}{dt}$, we need to relate the circuit time to boundary time: $\frac{d\tau}{dt} = \frac{2\pi}{\tilde\beta}$ where $\tilde T = \frac{1}{\tilde\beta}$ is the local energy scale. In vacuum AdS regime,
\begin{align*}
    &\tilde\beta = l\lambdabar(r) = 2\pi \frac{l^2}{r}\\
	&\frac{d\mathcal{C}}{dt} = \frac{d\mathcal{C}}{d\tau}\frac{d\tau}{dt} = s(\tau)\frac{2\pi}{\tilde\beta} = El\lambdabar \frac{2\pi}{l\lambdabar} = 2\pi E
\end{align*}
The complexity in the vacuum regime is given by
\begin{align}
\label{complexity_vacuum}
	\mathcal{C}(t) = \mathcal{C}_0+2\pi Et,\ \ \ \  \ t<\frac{l^2}{r_h}.
\end{align}

In near-Rindler regime, $\frac{d\tau}{dt} = \frac{2\pi}{\beta}$ and the complexity is given in \cite{Susskind:2014jwa}\cite{Roberts:2014isa}.
\begin{equation}
	\begin{split}
	\mathcal{C}(t) =\ &\mathcal{C}_0+\delta S+ S\log\left(1+\frac{\delta S}{S}e^{\frac{2\pi}{\beta}(t-\frac{l^2}{r_h})}\right)\\
	\approx\ &\mathcal{C}_0+\delta S+\begin{cases}
\delta S e^{\frac{2\pi}{\beta}(t-\frac{l^2}{r_h})} & \frac{l^2}{r_h}<t<\frac{\beta}{2\pi}\log\frac{S}{\delta S}\\
S\left(t-\frac{\beta}{2\pi}\log\frac{S}{\delta S}\right)	& t>\frac{\beta}{2\pi}\log\frac{S}{\delta S}
\end{cases}\label{complexity_Rindler}
\end{split}
\end{equation}.

In section \ref{size_momentum} we will compare the rate of increase of complexity in \eqref{complexity_vacuum} \eqref{complexity_Rindler} with the bulk radial momentum of an infalling object . 

\subsection{Complexity from holography}
Following complexity-volume duality\cite{Stanford:2014jda}, we can also estimate the change of complexity from the change of volume after a perturbation comes in\cite{Barbon:2019tuq}. The match of volume increase of the wormhole and complexity increase in epidemic model during and after scrambling regime was well checked\cite{Stanford:2014jda}\cite{Susskind:2014jwa}\cite{Roberts:2014isa}\cite{Brown:2015bva}\cite{Brown:2015lvg}\cite{Zhao:2017iul}. In this paper, we look at the increase of volume in the vacuum regime. We will use the relation
\begin{align}
\label{CV}
	\mathcal{C} = (D-2)\frac{\text{Vol}}{4G_N l}
\end{align}
where $D$ is bulk spacetime dimension. The dimensional dependent factor $D-2$ is chosen such that at late time, the complexity of a large black hole increases as $\mathcal{C}\sim 2\pi M t$ where $M$ is the mass of the black hole.

\subsubsection{Spherically symmetric perturbation}

We consider the following geometry. Before the particle comes in, we have vacuum AdS with metric
\begin{align*}
	ds^2 = -f(r)dt^2+\frac{dr^2}{f(r)}+r^2d\Omega_{D-2}^2
\end{align*}
where $f(r) = \frac{r^2}{l^2}+1$.
After the particle comes, the metric has the same form with
\begin{align*}
	\tilde f(r) = \frac{r^2}{l^2}+1-\frac{2m}{r^{D-3}} 
\end{align*}
where
\begin{align*}
	m = \frac{8\pi G_N E}{(D-2)\text{Vol}(\Omega_{D-2})}
\end{align*}

\begin{figure}[H] 
 \begin{center}                      
      \includegraphics[width=2.6in]{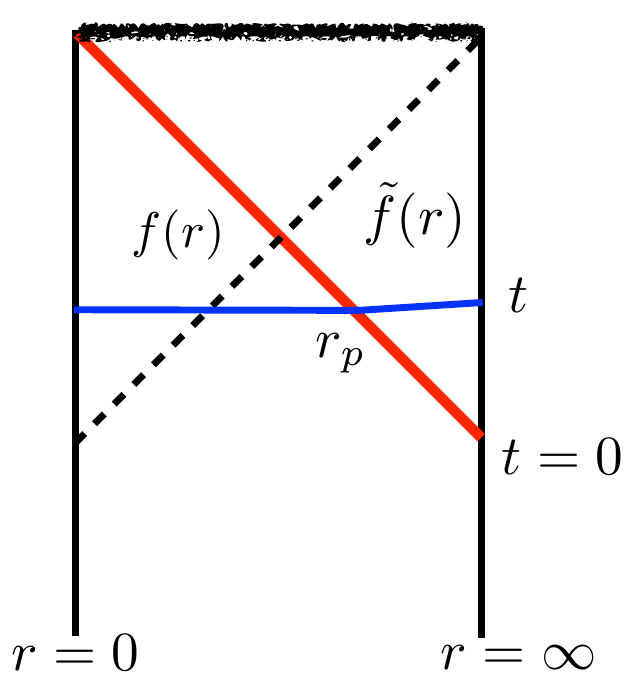}
      \caption{}
  \label{proof}
  \end{center}
\end{figure}

The two geometries are pasted along the red line in Figure \ref{proof}. We consider the slice at constant $t$ (blue line in Figure \ref{proof}). We put a cutoff at $r = r_c$ large. The volume of this slice at vacuum is given by
\begin{align*}
	\text{Vol}_{\text{vacuum}} = \text{Vol}(\Omega_{D-2})\int_0^{r_c}\frac{dr}{\sqrt{f(r)}}r^{D-2}
\end{align*}
Shortly after the perturbation falls in, we consider the volume of the same slice:
\begin{align*}
	\text{Vol}_{\text{perturbed}}(t) = \text{Vol}(\Omega_{D-2})\int_0^{r_p(t)}\frac{dr}{\sqrt{f(r)}}r^{D-2}+\text{Vol}(\Omega_{D-2})\int_{r_p(t)}^{R}\frac{dr}{\sqrt{\tilde f(r)}}r^{D-2}
\end{align*}
where $r_p(t)$ is the location of the object. 

For $D\geq 3$, we have
\begin{align*}
	&\frac{r^{D-2}}{\sqrt{\frac{r^2}{l^2}+1-\frac{2m}{r^{D-3}}}}-\frac{r^{D-2}}{\sqrt{\frac{r^2}{l^2}+1}}\\
	=\  &\frac{r^{D-2}}{\sqrt{\frac{r^2}{l^2}+1-\frac{2m}{r^{D-3}}}\sqrt{\frac{r^2}{l^2}+1}\left(\sqrt{\frac{r^2}{l^2}+1-\frac{2m}{r^{D-3}}}+\sqrt{\frac{r^2}{l^2}+1}\right)}\frac{2m}{r^{D-3}}\\
	=\ &\frac{l^3m}{r^2}\left(1+\mathcal{O}(\frac{l^2}{r^2})\right)
\end{align*}
We have
\begin{align*}
	\Delta \mathcal{C}(t) =\ & \frac{D-2}{4G_N l}\text{Vol}(\Omega_{D-2})\frac{l^3m}{r_p(t)} \\
	=\ & \frac{2\pi El^2}{r_p(t)}
\end{align*}
From \eqref{geodesic}, $\frac{l}{r_p}\approx\frac{t}{l}$, we have
\begin{align*}
	\Delta\mathcal{C}(t) = 2\pi E t
\end{align*}
This is the same as \eqref{complexity_vacuum} from circuit consideration.

\section{Complexity increase and the growth of momentum}
\label{size_momentum}
In $AdS_2$, the radial momentum of a particle is related to the rate of increase of the complexity\cite{Susskind:2019ddc}\cite{Lin:2019qwu}: $2\pi P =  \frac{d\mathcal{C}}{dt}$. We will generalize this relation to higher dimensions. We propose the following relation: 
\begin{align}
	\label{relation}
		\frac{\lambdabar}{2\pi}\frac{d\mathcal{C}}{dt} = 2\pi P
\end{align}
where $\lambdabar$ is the dimensionless boundary wavelength of the excitation. Note that $\frac{\lambdabar}{2\pi} = \frac{\beta}{2\pi l}$ at near Rindler region in $D\geq 3$.

The left side of \eqref{relation} is boundary quantity while the right side is bulk quantity. One can define $\lambdabar$ as the radius of gyration in \cite{Coleman:1977yb}: 
\begin{align*}
	\lambdabar = \frac{2\pi\bar R}{l}.
\end{align*}
The growth of $\lambdabar$ reflects that the perturbation is flowing from UV towards IR. From UV-IR connection, $\lambdabar$ is reflected in the radial location of the particle in the bulk. \\

In $AdS_2$ throat, $\frac{\lambdabar}{2\pi} \sim   1$ and \eqref{relation} reduces to \eqref{AdS2_3}. Let's look at other examples. 
\subsection{Vacuum AdS}
Consider vacuum AdS in coordinates \eqref{vaccum_AdS}. We consider the regime where $r\gg l$ so we can ignore $1$ in $\frac{r^2}{l^2}+1$. For a massless particle with energy $E$ falling in such a geometry, its radial momentum is given by
\begin{align*}
	P(t) = \frac{E}{\sqrt{f(r_p(t))}} = \frac{El}{r_p(t)}
\end{align*}
On the other hand, earlier we've seen that the complexity grow as $\mathcal{C}(t) = \mathcal{C}_0+2\pi E t$, $\frac{\lambdabar(t)}{2\pi} = \frac{t}{l}$, and $\frac{l}{r_p(t)} = \frac{t}{l}$. Combining these one can check that \eqref{relation} holds.

\subsection{Large black hole in AdS}
\begin{figure}[H] 
 \begin{center}                      
      \includegraphics[width=2.6in]{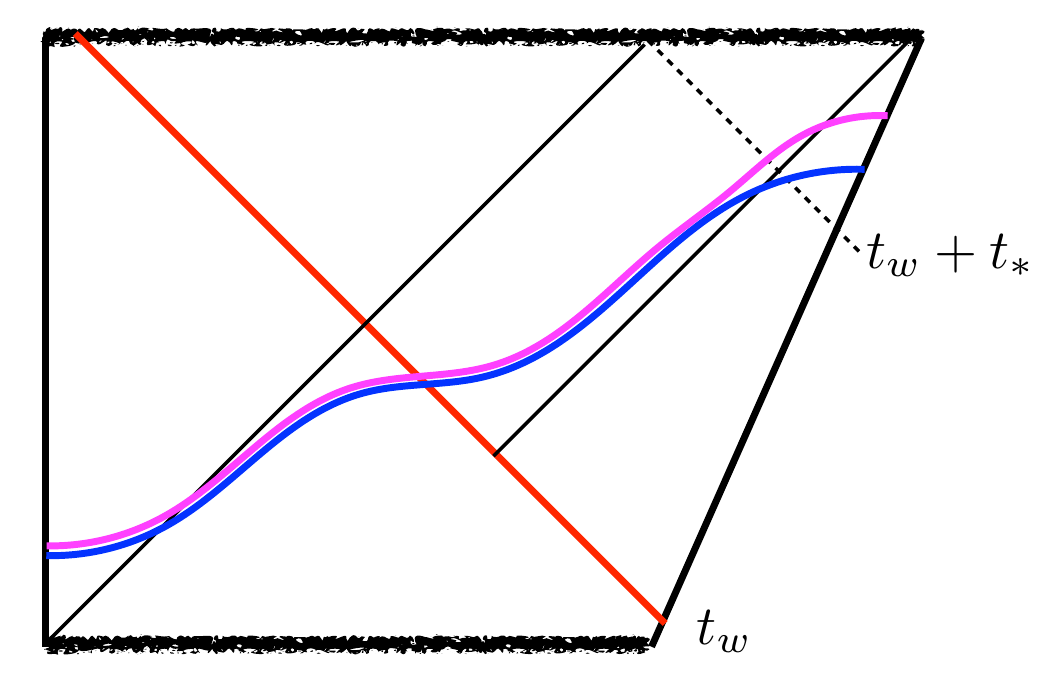}
      \caption{}
  \label{saturation}
  \end{center}
\end{figure}

The above discussion of size increase in vacuum AdS can be fit together with scrambling regime at finite temperature. Consider a BTZ black hole. 
From section \ref{finite_temperature},
\begin{align}
\label{wavelength_BTZ}
	\frac{\lambdabar}{2\pi} = \frac{l}{r} = \frac{l}{r_h}\tanh(\frac{r_h}{l^2}t)
\end{align}
The momentum of an infalling massless particle is given by
\begin{align}
\label{momentum_BTZ}
	2\pi P =\begin{cases}\frac{2\pi E}{\sqrt{f(r)}}=2\pi \frac{El}{r_h}\sinh(\frac{r_h}{l^2}t) & t<\frac{\beta}{2\pi}\log\frac{S}{\delta S}\\
	\frac{S}{l} & t>\frac{\beta}{2\pi}\log\frac{S}{\delta S}
\end{cases}
\end{align}
The saturation of momentum after scrambling time was discussed in \cite{Lin:2019qwu}. Roughly speaking, after scrambling time the backreaction becomes significant. Further time evolution will make the wormhole longer and does nothing to the particle deep in the interior. In Figure \ref{saturation}, we draw the maximal volume slice at two different times after scrambling time. We see that, as time increases, we simply add a piece to the outer edge of the wormhole. The intersection of the maximal volume slice with the infalling particle no longer changes.

If we plug \eqref{wavelength_BTZ}\eqref{momentum_BTZ} into \eqref{relation}, we get
\begin{align}
	\Delta \mathcal{C}(t) =\ & \begin{cases}
		\frac{2\pi El^2}{r_h}\sinh(\frac{r_h}{l^2}t) & t<t_*\\
		St &t>t_*
	\end{cases}\nonumber\\ 
	=\ &\begin{cases}
		2\pi Et & t<\frac{l^2}{r_h}\\
		\delta S e^{\frac{r_h}{l^2}t} & \frac{l^2}{r_h}<t<\frac{\beta}{2\pi}\log\frac{S}{\delta S}\\
		St & t>\frac{\beta}{2\pi}\log\frac{S}{\delta S}
	\end{cases}\label{complexity_circuit}
\end{align}
\eqref{complexity_circuit} is exactly what one expected about complexity from earlier discussions in \eqref{complexity_vacuum}\eqref{complexity_Rindler}.

\section{Conclusions}
\label{conclusion}
In this paper we studied the connections between operator size growth, complexity increase, and bulk radial momentum in spacetime dimension $D\geq 3$. We used a gluon-splitting model to study the operator growth and complexity increase in vacuum regime, and matched our result with CV calculation. We found a formula relating complexity increase to radial momentum that works in any spacetime dimension, from the time that the perturbation just comes in from the cutoff boundary, to after the scrambling time.

In the case of JT gravity and SYK, one can derive the relation between complexity and momentum from considerations of symmetry generators\cite{Lin:2019qwu}, or one can consider the Schwarzian boundary as a non-relativistic system and apply Newton's law\cite{Susskind:2019ddc}. 
On the other hand, so far our formula relating the complexity to radial momentum \eqref{relation} is empirical. It would be be nice give a first principle derivation of it or have something analogous to the non-relativistic system in the case of JT/SYK.

 \section*{Acknowledgments}

We thank Henry Lin, Alex Mousatov, and Steve Shenker for helpful discussions.
 Y.Z. thanks SITP for hospitality when part of this work was done. Y.Z. is supported by the Simons foundation through the It from Qubit Collaboration.


\end{document}